\documentstyle[12pt]{article}
\begin{document}
\title{TOV OR OV?, THE WHOLE STORY}
\author{L. Herrera$^1$\thanks{On leave from UCV, Caracas, Venezuela, e-mail: lherrera@usal.es}\\
{$^1$Instituto Universitario de F\'isica
Fundamental y Matem\'aticas},\\ {Universidad de Salamanca, Salamanca, Spain,}}
\maketitle

\begin{abstract}
In a recent manuscript published in the Arxives ( arXiv:1610.03049v1), it is claimed that it should be more appropriate to refer to  the equation of hydrostatic equilibrium in a static   spherically symmetric spacetime,  supported by an isotropic perfect fluid, as the ``Oppenheimer-Volkoff (OV) equation'', instead of the  ``Tolman-Oppenheimer-Volkoff (TOV)'', as is frequently  done. As we shall see in this brief note, such a claim is not supported by the existing bibliography.
\end{abstract}

\section{The bibliographic facts}

The argument put forward by Semiz \cite{1} to justify his claim, is based on a comparative study between  Tolman   \cite{a} and Oppenheimer, Volkoff \cite{b} papers. 

Here resides the root of his confusion. Indeed, the Tolman paper cited by Semiz \cite{a}, {\bf is not} about the TOV equation, but about the integration of the Einstein equations for spherically symmetric perfect fluids. Still worse, not a single word about  the former equation  appears in this reference. 

It is then obvious that if we restrict our analysis to \cite{a}, the ``T'' should be deleted from the TOV equation. However, the fact is that Tolman was  the first to obtain the TOV equation, but not in 1939, he did that  nine years before,  in 1930.

Indeed in \cite{2} the reader can check that Tolman's equation (22) is just the equation (9) in the Semiz manuscript (with an obvious change in notation). Furthermore, by evaluating  his equation in the Newtonian limit,  Tolman was able to identify it with the hydrostatic equilibrium equation, thereby providing the physical interpreation of this equation.

The same equation appears also  in \cite{3} (eq. 26), and  later on, well  before 1939, in \cite{4} (eqs.3.8 and 4.4). In this latter reference it is remarkable that Lemaitre deduced the TOV equation for the general case of an anisotropic fluid (principal stresses unequal), which of course becomes the usual TOV equation in the perfect fluid case, when both pressures are equal.

I am well aware of the fact that in references \cite{2,3,4} the TOV equation appears as  the equation (9) in \cite{1}, instead as in the more familiar form (13) (eq.(10) in \cite{b}).  The advantage  of the latter (from the point of view of its physical interpretation), with respect to the former, resides in the fact that the ``effective passive gravitational mass density'' ($p+\rho$), and the ``effective active  gravitational mass density'' ($4\pi pr^3+u$) where $u$ denotes the mass function, appear explicitly. This could be a reason to add OV to the T. However the obtention of (13) from (9) is rather elementary, and by no means, justifies to delete the T in TOV. At any rate, if a change should be done to the name of the equation, it should be to TLOV (where L stands for Lemaitre).

\end{document}